\documentclass[prd,amsfonts,noshowpacs,nofootinbib]{revtex4}
\usepackage{amsmath}
\usepackage {graphicx}

\begin{document}

\begin{abstract}
We compute the tree-level connected four-point function of the primordial curvature perturbation for a fairly general minimally coupled single field inflationary model, where the inflaton's Lagrangian is a general function of the scalar field and its first derivatives. This model includes K-inflation and DBI-inflation as particular cases. We show that, at the leading order in the slow-roll expansion and in the small sound speed limit, there are two important tree-level diagrams for the trispectrum. One is a diagram where a scalar mode is exchanged and the other is a diagram where the interaction occurs at a point, i.e. a contact interaction diagram. The scalar exchange contribution is comparable to the contact interaction contribution. For the DBI-inflation model, in the so-called equilateral configuration, the scalar exchange trispectrum is maximized when the angles between the four momentum vectors are equal and in this case the amplitude of the trispectrum from the scalar exchange is one order of magnitude higher than the contact interaction trispectrum.
\end{abstract}

\title{On the full trispectrum in single field DBI-inflation}

\author{Frederico Arroja\footnote{arrojaf@yukawa.kyoto-u.ac.jp}$\flat$}
\author{Shuntaro Mizuno\footnote{shuntaro.mizuno@nottingham.ac.uk}$\sharp$}
\author{Kazuya Koyama\footnote{Kazuya.Koyama@port.ac.uk}$\natural$}
\author{Takahiro Tanaka\footnote{tanaka@yukawa.kyoto-u.ac.jp}$\flat$}
\affiliation{
$\flat$Yukawa Institute for Theoretical Physics, Kyoto University, Kyoto 606-8502, Japan.
\\
$\sharp$School of Physics and Astronomy, University of Nottingham, University Park, Nottingham NG7 2RD, UK;
Research Center for the Early Universe (RESCEU), Graduate School of Science, The University of Tokyo, Tokyo 113-0033, Japan.
\\
$\natural$Institute of Cosmology and Gravitation, University of Portsmouth, Portsmouth PO1 2EG, UK.
}

\date{\today}
\maketitle


\section{\label{sec:INTRO}Introduction}

Precise measurements of the cosmic microwave background (CMB)
anisotropies such as those obtained by the WMAP satellite \cite{WMAP} provide
valuable information on the very early universe. Any theoretical
model that attempts to explain the evolution of the universe before the
big bang nucleosynthesis will also have to explain the observed CMB
anisotropies. The power spectrum of these primordial anisotropies is
nearly scale invariant \cite{Komatsu:2008hk} and it contains
almost all the information on primordial perturbations. However,
a small amount of non-Gaussianity is still allowed by the data
\cite{Komatsu:2008hk,Smith:2009jr}.
The information contained in this non-Gaussian component
will contribute to a huge advance in our understanding of
the very early universe. For example, the simplest and most popular
inflationary model based on a single field with a canonical kinetic term
satisfying slow-roll conditions and standard initial conditions predicts
that the level of primordial non-Gaussianit is actually unobservably small
\cite{Maldacena:2002vr}, even with future CMB experiments like the Planck
satellite \cite{PLANCK}.

Recently however, there are some hints of non-Gaussianity \cite{Yadav:2007yy,Komatsu:2008hk,Smith:2009jr}
in the CMB anisotropies. These are one (or so) standard deviation hints but they motivate tremendous
efforts from the theoretical cosmology community to create inflationary models that can produce large levels of non-Gaussianity both seen in the bispectrum and in the trispectrum. In the literature, there is a myriad \cite{Linde:1996gt,Bartolo:2001cw,Bernardeau:2002jy,Bernardeau:2002jf,Dvali:2003em,Creminelli:2003iq,Alishahiha:2004eh,Gruzinov:2004jx,Enqvist:2004ey,Jokinen:2005by,Lyth:2005qk,Salem:2005nd,Seery:2006js,
Sasaki:2006kq,Malik:2006pm,Barnaby:2006cq,Alabidi:2006wa,Chen:2006nt,Huang:2006eh,Chen:2006xjb,Alabidi:2006hg,Byrnes:2006vq,Suyama:2007bg,Arroja:2008ga,Arroja:2008yy,Langlois:2008wt,Langlois:2008qf,
Sasaki:2008uc,Byrnes:2008wi,Byrnes:2008zy,Dutta:2008if,Naruko:2008sq,Suyama:2008nt,Gao:2008dt,Cogollo:2008bi,Rodriguez:2008hy,Ichikawa:2008iq,Byrnes:2008zz,Li:2008fma,Langlois:2008vk,Hikage:2008sk,
Kawasaki:2008sn,Gao:2009gd,Cai:2009hw,Langlois:2009ej,Gao:2009bx,Huang:2009xa} of models that can produce sizeable non-Gaussianity. They basically relax one (or more) of the following conditions: single field, slow-roll, standard kinetic energy and standard initial conditions.

In the present work, we will consider a single field model with a non-canonical kinetic term. It has been shown \cite{Alishahiha:2004eh,Gruzinov:2004jx,Chen:2006nt,Huang:2006eh} that this kind of models predicts large bispectrum and also large trispectrum of the primordial curvature perturbation. In particular the DBI-inflation model \cite{Silverstein:2003hf}, which is constructed within string theory, is such an example. For current and stringent observational constraints and consequences of DBI-inflation see \cite{Kecskemeti:2006cg,Lidsey:2006ia,Baumann:2006cd,Bean:2007hc,Lidsey:2007gq,Peiris:2007gz,Kobayashi:2007hm,Lorenz:2007ze,Bird:2009pq}.

We will focus our attention on the trispectrum of the primordial curvature perturbation. The current observational bound \cite{Boubekeur:2005fj,Alabidi:2005qi} on the trispectrum is rather weak $|\tau_{NL}|<10^8$ , where $\tau_{NL}$ parameterizes the size of the trispectrum, but in the near future with the Planck satellite it will improve to $|\tau_{NL}|\sim560$ \cite{Kogo:2006kh}. We should note that these bounds depend on the shape of the trispectrum \cite{Babich:2004gb} and the values quoted here can be applied only for local models of
non-Gaussianity. Thus it is important to calculate the exact form of the trispectrum in the DBI-inflation model
and construct an estimator to measure the trispectrum of the CMB anisotropies.

In \cite{Seery:2006vu}, Seery \emph{et al.} calculated the contact interaction trispectrum for slow-roll multiple fields inflation but with standard kinetic terms. They showed that at horizon crossing the contact interaction trispectrum is too small to be observed. They found $\tau_{NL}^{CI}\sim r$, where $r$ is the tensor to scalar ratio, that is currently constrained to the range $r<0.22$ at 95\% confidence level \cite{Komatsu:2008hk}. Recently, Seery \emph{et al.} \cite{Seery:2008ax}, computed the graviton exchange trispectrum and they showed that the above conclusion does not change, i.e. the total non-linearity parameter $\tau_{NL}$ coming from both the contact interaction and the graviton exchange trispectrum is still unobservably small and of order $r$. Also recently, Gao and Hu \cite{Gao:2009gd} computed the leading order trispectrum from entropy perturbations in multifield DBI-inflation model.

In \cite{Huang:2006eh} (see also \cite{Arroja:2008ga}), the authors computed the trispectrum at leading order in slow-roll and in the small sound speed limit for a fairly general inflationary model, where the inflaton's Lagrangian is a general function of the scalar field and its first derivatives. They only calculated the contact interaction trispectrum and they argue that for certain models (DBI-inflation inclusive) the trispectrum can reach values in the observable range of the Planck satellite.

In the present work, we will calculate the scalar exchange trispectrum for the same model. We will show that the scalar exchange trispectrum is of the same importance as the contact interaction trispectrum. This completes the work of \cite{Huang:2006eh} (see also \cite{Arroja:2008ga}) and gives the final answer that should be constrained when comparing the trispectrum for models with large non-Gaussianity with observations.

The rest of this paper is organized as follows. In the next section, we shall introduce the model and some basic notations. In section \ref{sec:Perturbations}, we will consider non-linear perturbations in the action up to quartic order. In subsection \ref{subsec:OM}, we will present a simple argument to prove that the amplitude
of the scalar exchange trispectrum is of the same order of magnitude as the contact interaction trispectrum. In subsection \ref{subsec:CI}, we will calculate the trispectrum of the primordial curvature perturbation at leading order in slow-roll and in the small sound speed limit when only contact interactions are taken into account. In subsection \ref{subsec:SE}, we shall calculate the scalar exchange trispectrum. At the end of this subsection, we present for the first time the complete trispectrum at leading order in the previously mentioned approximations.
Section \ref{sec:Shape} is devoted to the study of the shape of the trispectrum in the so-called equilateral configuration for the DBI-inflation model. We will also calculate the non-linearity parameter $\tau_{NL}$ for different models. The conclusions of this work are described in section \ref{sec:conclusion}.


\section{\label{sec:MODEL}The model}
In this section, we will introduce the inflationary model under study. We shall present the background equations of motion and define the slow-variation parameters.
At the end of the section, we show the results for the primordial power spectrum of the curvature perturbation and the spectral index \cite{Garriga:1999vw}.

We will consider a fairly general class of models
described by the following action
\begin{equation}
S=\frac{1}{2}\int d^4x\sqrt{-g}\left[M^2_{Pl}R+2P(X,\phi)\right],
\label{action}
\end{equation}
where $\phi$ is the inflaton field, $M_{Pl}$ is the reduced Planck mass, $R$  is the Ricci scalar and
$X\equiv-(1/2)g^{\mu\nu}\partial_\mu\phi\partial_\nu\phi$, where $g_{\mu\nu}$ is the metric tensor.
We label the inflaton's Lagrangian by $P$ and we assume that it is
a well behaved function of two variables, the inflaton field and $X$. Throughout this work, we use a system of units where the Planck constant $\hbar$, the speed of light $c$ and the reduced Planck mass $M_{Pl}$ are
set to unity. This general Lagrangian includes as particular cases the DBI-inflation model
\cite{Alishahiha:2004eh,Silverstein:2003hf} and the K-inflation model
\cite{ArmendarizPicon:1999rj}.
In the DBI-inflation model $P(X,\phi)$ is given by
\begin{equation}
P(X,\phi) = -f(\phi)^{-1} \sqrt{1- 2 f(\phi) X} -V(\phi),
\end{equation}
where $f(\phi)$ and $V(\phi)$ are functions of the scalar field determined by string theory
configurations. At the background level, we assume that our Universe is well described by a flat,
homogeneous and isotropic
Friedmann-Robertson-Walker universe given by the line element
\begin{equation}
ds^2=-dt^2+a^2(t)\delta_{ij}dx^idx^j, \label{FRW}
\end{equation}
where $a(t)$ is the scale factor. The Friedmann equation and the
continuity equation read
\begin{equation}
3H^2=E, \label{EinsteinEq}
\end{equation}
\begin{equation}
\dot{E}=-3H\left(E+P\right), \label{continuity}
\end{equation}
where dot denotes derivative with respect to cosmic time, the Hubble rate is $H=\dot{a}/a$, $E$ is the energy of the
inflaton and it is given by
\begin{equation}
E=2XP_{,X}-P
,\label{energy}
\end{equation}
where $P_{,X}$ denotes the derivative of $P$ with respect to $X$.
It was shown in \cite{Garriga:1999vw} (see also \cite{Christopherson:2008ry}) that for this model the
speed of propagation of scalar perturbations (``speed of sound")
is $c_s$ given by
\begin{equation}
c_s^2=\frac{P_{,X}}{E_{,X}}=\frac{P_{,X}}{P_{,X}+2XP_{,XX}}. \label{sound
speed}
\end{equation}
We define the slow variation parameters, analogues of the
slow-roll parameters, as:
\begin{equation}
\epsilon=-\frac{\dot{H}}{H^2}=\frac{XP_{,X}}{H^2}, \quad
\xi=\frac{\dot{\epsilon}}{\epsilon H}, \quad
s=\frac{\dot{c_s}}{c_sH}.
\end{equation}
We should note that these slow variation parameters are more
general than the usual slow-roll parameters and that the smallness
of these parameters does not imply that the field in rolling
slowly. We assume that the rate of change of the speed of sound is
small (as described by $s$) but $c_s$ is otherwise free to change
between zero and one.

The power spectrum of the primordial quantum
fluctuation was first derived in \cite{Garriga:1999vw} and reads
\begin{equation}
{\cal P}_\zeta(k)=\frac{1}{36\pi^2}\frac{E^2}{E+P}=\frac{1}{8\pi^2}\frac{H^2}{c_s\epsilon},
\label{PowerSpectrum}
\end{equation}
where it should be evaluated at the time of horizon crossing
${c_s}_*k=a_*H_*$. The spectral index is
\begin{equation}
n_s-1=\frac{d\ln {\cal P}_\zeta(k)}{d\ln k}=-2\epsilon-\xi-s.
\label{SpectralIndex}
\end{equation}
WMAP observations of the perturbations in the CMB tell us that
the previous power spectrum is almost scale invariant therefore
implying that the three slow variation parameters should be small
at horizon crossing, roughly of order $10^{-2}$.

\section{\label{sec:Perturbations}Non-linear perturbations}

The main goal of this section is to calculate the total connected trispectrum of the primordial curvature perturbation for the model (\ref{action}), at leading order in the slow-roll expansion and in the limit of small sound speed.
In the next subsection, with very simple arguments, we will estimate the order of magnitude of the trispectrum coming from the scalar exchange and compare it with the result of the trispectrum coming from the contact interaction \cite{Huang:2006eh} (see also \cite{Arroja:2008ga}).
In subsection \ref{subsec:CI}, we review the calculation of the contact interaction trispectrum. Finally, in subsection \ref{subsec:SE}, we calculate the scalar exchange trispectrum. The total trispectrum will be the sum of these contributions.

Non-linear perturbations of the model (\ref{action}) have been considered by several authors before. In \cite{Gruzinov:2004jx}, Gruzinov obtained the bispectrum using the simple method of expanding the action of the inflaton only. This method gives the leading order bispectrum in slow-roll and in the small sound speed limit. In \cite{Chen:2006nt}, the authors performed a detailed analysis of the bispectrum produced in the present model including also next order terms in slow-roll and small sound speed expansion. The trispectrum has also been studied in the literature. In particular, Chen \emph{et al.} \cite{Huang:2006eh} (see also \cite{Arroja:2008ga}) calculated the trispectrum from the contact interaction.

The Lagrangian density of the inflaton expanded up to fourth order in perturbations (at leading order in slow-roll and in the small sound speed limit) can be founded in \cite{Huang:2006eh}, and it reads
\begin{equation}
\mathcal{L}^{(2)}=\frac{a^3P_{,X}}{2c_s^2}\dot{\alpha}^2 -\frac{aP_{,X}}{2}(\partial\alpha)^2 ,\label{2action}
\end{equation}

\begin{equation}
\mathcal{L}^{(3)}=\left(P_{,XX}\frac{\dot\phi}{2}+P_{,XXX}\frac{\dot\phi^3}{6}\right)a^3\dot{\alpha}^3-P_{,XX}\frac{\dot\phi}{2}a\dot{\alpha}(\partial\alpha)^2 ,
\label{3action}
\end{equation}

\begin{equation}
\mathcal{L}^{(4)}=\frac{a^3}{4}\bigg[\left(\frac{\dot\phi^4}{6}P_{,4X}+\dot\phi^2P_{,XXX}+\frac{1}{2}P_{,XX}\right)\dot{\alpha}^4-a^{-2}\dot{\alpha}^2(\partial\alpha)^2\left(\dot\phi^2P_{,XXX}+P_{,XX}\right)+\frac{1}{2}a^{-4}P_{,XX}(\partial\alpha)^4\bigg] ,\label{4action}
\end{equation}
where $\alpha$ denotes the field perturbation $\delta\phi(t,\mathbf{x})$.

In the next subsections, we will use the so-called ``interaction picture formalism" \cite{Weinberg:2005vy} to calculate the four-point quantum correlation function. In this formalism one needs to know the interaction Hamiltonian.
The Hamiltonian density $\mathcal{H}$ can be found using Eqs. (\ref{2action}), (\ref{3action}), (\ref{4action}) and
\begin{equation}
\mathcal{H}=\pi\dot\alpha-\mathcal{L},
\end{equation}
where $\mathcal{L}$ is the Lagrangian density and $\pi$ is the momentum density, defined by $\pi=\partial\mathcal{L}/\partial\dot\alpha$.
The third order Hamiltonian density is $\mathcal{H}^{(3)}=-\mathcal{L}^{(3)}$ and the fourth order Hamiltonian density is \cite{Huang:2006eh}
\begin{equation}
\mathcal{H}^{(4)}=\beta_1a^3\dot{\alpha}^4+\beta_2a\dot{\alpha}^2(\partial\alpha)^2+\beta_3a^{-1}(\partial\alpha)^4 ,\label{4H}
\end{equation}
where we have used $\dot\alpha=\partial\mathcal{H}_0/\partial\pi$ to express $\pi$ in terms of $\dot\alpha$, and $\mathcal{H}_0$ denotes the kinematic Hamiltonian density that is quadratic in $\pi$ and $\alpha$. Finally to obtain the interaction Hamiltonian from the (\ref{4H}), the variables $\dot\alpha$ and $\alpha$ should be replaced with their interaction picture counterparts $\dot\alpha_I$ and $\alpha_I$.

The constants $\beta_i$ are defined as
\begin{eqnarray}
\beta_1&=&P_{,XX}\left(1-\frac{9}{8}c_s^2\right)-\dot\phi^2P_{,XXX}\left(1-\frac{3}{4}c_s^2\right)+\frac{1}{8}\frac{\dot\phi^6c_s^2}{P_{,X}}P_{,XXX}^2-\frac{1}{24}\dot\phi^4P_{,4X},
\nonumber\\
\beta_2&=&-\frac{1}{2}P_{,XX}\left(1-\frac{3}{2}c_s^2\right)+\frac{1}{4}\dot\phi^2c_s^2P_{,XXX},
\nonumber\\
\beta_3&=&-\frac{c_s^2}{8}P_{,XX}.
\end{eqnarray}
For DBI inflation, at leading order in the sound speed, the constants are
\begin{equation}
\beta_1=\frac{1}{2c_s^7\dot\phi^2}, \quad \beta_2=\frac{1}{4c_s^3\dot\phi^2}, \quad \beta_3=-\frac{1}{8c_s\dot\phi^2}.
\end{equation}

From the second-order action, we can solve for the perturbation and
quantize it according to the standard procedures of quantum field
theory:
\begin{eqnarray}
\alpha(\eta,\textbf{x})&=&\frac{1}{(2\pi)^3}\int d^3k
[u(\eta,{\textbf{k}})a(\textbf{k})
+u^*(\eta,{-\textbf{k}})a^{\dagger}(-\textbf{k})]e^{i\textbf{k}\cdot\textbf{x}}~,
\end{eqnarray}
where
$u(\eta,{\textbf{k}})=\frac{H}{\sqrt{2k^3}}\frac{1}{\sqrt{P_{,X}c_s}}(1+ikc_s\eta)e^{-ikc_s\eta}$
is the solution of the quadratic Lagrangian, and
$\eta=-\frac{1}{aH}$ is the conformal time.
The commutation relations are given by
\begin{eqnarray}
[a(\mathbf{k}_1),a(\mathbf{k}_2)]=[a^\dag(\mathbf{k}_1),a^\dag(\mathbf{k}_2)]=0, \quad [a(\mathbf{k}_1),a^\dag(\mathbf{k}_2)]=(2\pi)^3\delta^{(3)}(\mathbf{k}_1-\mathbf{k}_2).
\end{eqnarray}

\subsection{\label{subsec:OM}Estimate of the amplitude of the total trispectrum}

In this subsection, we shall present a very simple argument to prove that the trispectrum coming from the contact interaction diagram is of the same order as the trispectrum coming from the scalar exchange diagram. Therefore the total trispectrum is the sum of these two contributions. For simplicity, we will consider the DBI-inflation case but the conclusion is valid for generic models of the form (\ref{action}).

\begin{figure}[t]
\centering
 \scalebox{.8}
 {\rotatebox{0}{
    \includegraphics*{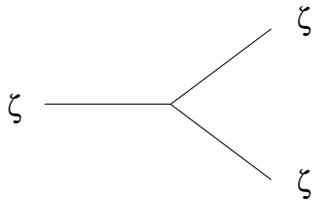}
                 }
 }
\caption{Bispectrum diagram}\label{BispectrumDiagram}
\end{figure}

\begin{figure}[t]
\centering
 \scalebox{.8}
 {\rotatebox{0}{
    \includegraphics*{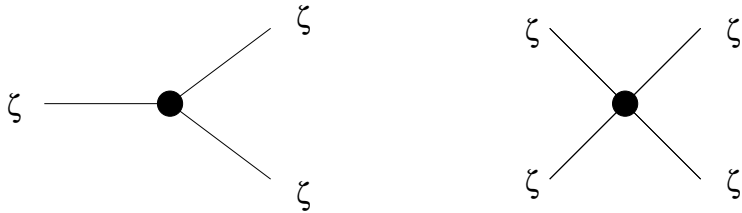}
                 }
 }
\caption{On the left: the third order vertex. On the right: the fourth order vertex }\label{Vertices}
\end{figure}

\begin{figure}[t]
\centering
 \scalebox{.8}
 {\rotatebox{0}{
    \includegraphics*{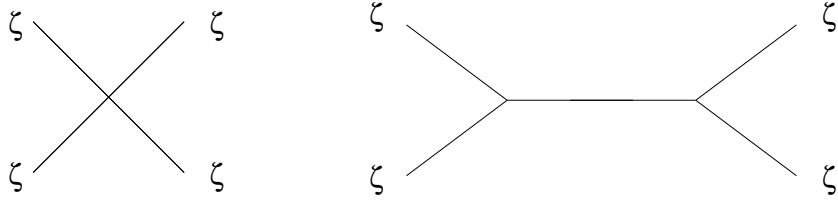}
                 }
 }
\caption{Trispectrum diagrams. On the left: the contact interaction. On the right: The interaction via exchange of a scalar.}\label{TrispectrumDiagrams}
\end{figure}

Let us start by drawing the tree-level diagram that contributes to the bispectrum. It can be found in Fig. \ref{BispectrumDiagram}, and it is composed of three external lines labeled by $\zeta$ and it has three scalar propagators ${\cal P}_\zeta\sim H^2\epsilon^{-1}c_s^{-1}$, and one third order vertex $I_3$ as depicted in the l.h.s. of Fig. \ref{Vertices}. The bispectrum coming from this diagram is then $B_\zeta\sim {\cal P}_\zeta^3I_3$.
The third order vertex is given by $I_3\sim H^{-2}\epsilon c_s^{-1}$ as is easily seen from the third
order Hamiltonian \cite{Chen:2006nt}, thus the DBI-bispectrum scales like $B_\zeta\sim c_s^{-2} {\cal P}_\zeta^2$ (or equivalently the non-linearity parameter $f_{NL}$ scales like $f_{NL}\sim c_s^{-2}$).
For the trispectrum, there are two relevant tree-level diagrams, Fig. \ref{TrispectrumDiagrams}. If we denote the fourth order vertex on the r.h.s. of Fig. \ref{Vertices} by $I_4$ then the diagram on the l.h.s. of Fig. \ref{TrispectrumDiagrams} gives a contribution for the trispectrum as $T^{CI}\sim {\cal P}_\zeta^4I_4$ and the diagram on the r.h.s. of the same figure gives $T^{SE}\sim {\cal P}_\zeta^5I_3^2$. From the fourth order Hamiltonian,
the fourth order vertex is estimated as $I_4\sim H^{-2}\epsilon c_s^{-3}$ thus the trispectrum from the
contact interaction is given by $T^{CI}\sim c_s^{-4} {\cal P}_\zeta^3$ \cite{Huang:2006eh}. The magnitude of $T^{SE}$ is easily calculated as $T^{SE}\sim c_s^{-4} {\cal P}_\zeta^3$ which has the same magnitude as $T^{CI}$. Therefore for consistency the total trispectrum is the sum of $T^{CI}$ and $T^{SE}$.

Recently, \cite{Seery:2008ax} showed that in the case of the standard kinetic term inflation, i.e. when $P$ is $P=X-V(\phi)$ ($V(\phi)$ is the potential), the diagram where we have an exchange of a graviton can in fact dominate the contact interaction diagram. In this model, the contribution from the scalar exchange diagram is negligible in comparison with both the contact interaction and the graviton exchange. They showed that the contact interaction trispectrum is $T^{CI}\sim\epsilon {\cal P}_\zeta^3$, the graviton exchange trispectrum is $T^{GE}\sim\epsilon {\cal P}_\zeta^3$ and the scalar exchange trispectrum is $T^{SE}\sim \epsilon^2{\cal P}_\zeta^3$. The third order scalar vertex is $I_{\zeta^3}\sim H^{-2}\epsilon^2$ and the fourth order vertex is $I_{\zeta^4}\sim H^{-2}\epsilon^2$.

We have checked that in the case of non-standard kinetic terms the graviton exchange diagram gives a contribution to the trispectrum that is suppressed by $\epsilon$ and $c_s$ with respect to the leading order and we neglect this contribution. This fact can be easily understood if one uses the main result of \cite{Seery:2008ax} and one recalls Maldacena's result \cite{Maldacena:2002vr} that the third order vertex of two-scalars-one-graviton is $I_{\gamma\zeta^2}\sim H^{-2}\epsilon c_s$.

In the next subsections, we will derive the momentum dependence of the trispectrum from the contact interaction and the scalar exchange.


\subsection{\label{subsec:CI}The trispectrum from the contact interaction}

In this subsection, we shall make use of the so-called ``interaction picture formalism" \cite{Weinberg:2005vy} to calculate the four-point function of the field perturbation at horizon crossing coming from a diagram like the one on the l.h.s. of Fig. \ref{TrispectrumDiagrams}. The result can also be found in Chen \emph{et al.} \cite{Huang:2006eh} (see also \cite{Arroja:2008ga}). These results are at leading order in slow-roll and in the limit of small sound speed.
For the contact interaction diagram, the four-point function is
\begin{eqnarray}
\langle\Omega|\delta\phi(0,\mathbf{k}_1)\delta\phi(0,\mathbf{k}_2)&&\!\!\!\!\!\!\!\delta\phi(0,\mathbf{k}_3)\delta\phi(0,\mathbf{k}_4)|\Omega\rangle^{CI}
\nonumber\\
&&= -i\int_{-\infty}^0 d\eta \langle 0|\left[\delta\phi_I(0,\mathbf{k}_1)\delta\phi_I(0,\mathbf{k}_2)\delta\phi_I(0,\mathbf{k}_3)\delta\phi_I(0,\mathbf{k}_4),H_I^{(4)}(\eta)\right]|0\rangle,
\end{eqnarray}
where $H_I(\eta)^{(4)}$ is the fourth order interaction Hamiltonian given by $H_I^{(4)}=\int d^3x\mathcal{H}_I^{(4)}$.
It can be found from Eq. (\ref{4H}) and reads
\begin{eqnarray}
H_I^{(4)}(\eta)=\int d^3x\left[\beta_1\delta\phi_I'^4+\beta_2\delta\phi_I'^2\left(\partial\delta\phi_I\right)^2+\beta_3\left(\partial\delta\phi_I\right)^4\right],
\end{eqnarray}
where prime denotes derivative with respect to conformal time $\eta$.

For a generic K-inflation model, we obtain
\begin{eqnarray}
\langle\Omega|&&\!\!\!\!\!\!\!\!\delta\phi(0,\mathbf{k}_1)\delta\phi(0,\mathbf{k}_2)
\delta\phi(0,\mathbf{k}_3)\delta\phi(0,\mathbf{k}_4)|\Omega\rangle^{CI}
=-(2\pi)^3\delta^{(3)}(\mathbf{K})\times
\nonumber\\
&&
\Bigg[
1152\beta_1N^8c_s^3\frac{1}{\left(\sum_{i=1}^4k_i\right)^5\Pi_{i=1}^4k_i}
\nonumber\\
&&
\quad+4\beta_2N^8c_s
\frac{k_1^2k_2^2 (\mathbf{k}_3\cdot\mathbf{k}_4)}{\left(\sum_{i=1}^4k_i\right)^3\Pi_{i=1}^4k_i^3}
\left(1+3\frac{k_3+k_4}{\sum_{i=1}^4k_i}+12\frac{k_3k_4}
{\left(\sum_{i=1}^4k_i\right)^2}\right)+\mathrm{permutations}
\nonumber\\
&&
\quad+32\beta_3\frac{N^8}{c_s}
\frac{(\mathbf{k}_1\cdot\mathbf{k}_2)(\mathbf{k}_3\cdot\mathbf{k}_4)+(\mathbf{k}_1\cdot\mathbf{k}_3)
(\mathbf{k}_2\cdot\mathbf{k}_4)+(\mathbf{k}_1\cdot\mathbf{k}_4)(\mathbf{k}_2\cdot\mathbf{k}_3)}{\sum_{i=1}^4k_i\,\Pi_{i=1}^4k_i^3} \nonumber\\
&&
\qquad\times\left(1+\frac{\sum_{i<j}k_ik_j}{\left(\sum_{i=1}^4k_i\right)^2}
+3\frac{\Pi_{i=1}^4k_i}{\left(\sum_{i=1}^4k_i\right)^3}\sum_{i=1}^4
\frac{1}{k_i}+12\frac{\Pi_{i=1}^4k_i}{\left(\sum_{i=1}^4k_i\right)^4}\right)\Bigg],
\label{CITrispectrumPHI}
\end{eqnarray}
where the total momentum $\mathbf{K}$ is defined by $\mathbf{K}=\mathbf{k}_1+\mathbf{k}_2+\mathbf{k}_3+\mathbf{k}_4$, the constant $N$ is defined by $N=H/\sqrt{2P_{,X}c_s}$ and ``permutations" denote the other twenty three terms that result from the permutations of $\{k_1,k_2,k_3,k_4\}$ in the preceding term. The three dimensional Dirac delta function of $\mathbf{K}$ ensures momentum conservation and implies that the four momentum vectors $\mathbf{k}_i$, where $i=1,\ldots,4$, form a quadrilateral.

For DBI-inflation, in the small sound speed limit, the terms proportional to $\beta_2$ and $\beta_3$ are sub-leading with respect to the term proportional to $\beta_1$.
For K-inflation models, this is not necessarily the case and this fact can be used to distinguish some K-inflation models from DBI-inflation \cite{Huang:2006eh}.

To obtain the four-point correlation function of the curvature perturbation at some time after horizon crossing we can use the linear relation $\zeta=-(H/\dot\phi)\delta\phi$ because
the higher order terms in this relation only generate sub-leading corrections to this result and we ignore them. Finally, we obtain
\begin{eqnarray}
\langle\Omega|\zeta(0,\mathbf{k}_1)\zeta(0,\mathbf{k}_2)\zeta(0,\mathbf{k}_3)\zeta(0,\mathbf{k}_4)|\Omega\rangle^{CI}=\frac{H^4}{\dot\phi^4}\langle\Omega|\delta\phi(0,\mathbf{k}_1)\delta\phi(0,\mathbf{k}_2)\delta\phi(0,\mathbf{k}_3)\delta\phi(0,\mathbf{k}_4)|\Omega\rangle^{CI}.
\end{eqnarray}

\subsection{\label{subsec:SE}The trispectrum from the scalar exchange interaction}
Within the ``interaction picture" formalism, to calculate the four-point function resulting from a correlation established via the exchange of a scalar mode as depicted in the r.h.s. of Fig. \ref{TrispectrumDiagrams} one needs to evaluate the following time integrals
\begin{eqnarray}
\langle\Omega|\delta\phi(0,\mathbf{k}_1)\delta\phi(0,\mathbf{k}_2)&&\!\!\!\!\!\!\!\delta\phi(0,\mathbf{k}_3)\delta\phi(0,\mathbf{k}_4)|\Omega\rangle^{SE}
\nonumber\\
&&= -\int_{-\infty}^0 d\eta\int_{-\infty}^\eta d\tilde\eta \langle
 0|\left[\left[\delta\phi_I(0,\mathbf{k}_1)\delta\phi_I(0,\mathbf{k}_2)\delta\phi_I(0,\mathbf{k}_3)\delta\phi_I(0,\mathbf{k}_4),H_I^{(3)}(\eta)\right],H_I^{(3)}(\tilde\eta)\right]|0\rangle,\nonumber\\ \label{CIintegral}
\end{eqnarray}
where $H_I^{(3)}(\eta)$ is the third order interaction Hamiltonian given by $H_I^{(3)}=\int d^3x\mathcal{H}_I^{(3)}=-\int d^3x\mathcal{L}_I^{(3)}$.
We will use
\begin{eqnarray}
H_I^{(3)}(\eta)=\int d^3x\left[Aa\delta\phi_I'^3+Ba\delta\phi_I'\left(\partial\delta\phi_I\right)^2\right],
\end{eqnarray}
where the constants $A$ and $B$ are defined as
\begin{equation}
A=-\frac{\dot\phi}{2}\left(P_{,XX}+\frac{1}{3}\dot\phi^2P_{,XXX}\right), \quad B=\frac{\dot\phi}{2}P_{,XX}.
\end{equation}
For DBI inflation they are given by
\begin{equation}
A=-\frac{1}{2\dot\phi c_s^5}, \quad B=\frac{1}{2\dot\phi c_s^3}.
\end{equation}

The calculation of the integral in (\ref{CIintegral}) is rather long but relatively straightforward (however see \cite{Adshead:2009cb}). The total four-point function of the field perturbations is
\begin{eqnarray}
\langle\Omega|&&\!\!\!\!\!\!\!\!\delta\phi(0,\mathbf{k}_1)\delta\phi(0,\mathbf{k}_2)\delta\phi(0,\mathbf{k}_3)\delta\phi(0,\mathbf{k}_4)|\Omega\rangle^{SE}
=(2\pi)^3\delta^{(3)}(\mathbf{K})\frac{2N^4}{(k_1k_2k_3k_4)^\frac{3}{2}}\times
\nonumber\\
&&\Bigg[
-9A^2\bigg(\mathcal{F}_1(k_1,k_2,-k_{12},k_3,k_4,k_{12})-\mathcal{F}_1(-k_1,-k_2,-k_{12},k_3,k_4,k_{12})\bigg)
\nonumber\\&&
\,\,\,\,\,
+AB
\Bigg(
      3(\mathbf{k}_3\cdot\mathbf{k}_4)\bigg(\mathcal{F}_3(k_1,k_2,-k_{12},k_{12},k_3,k_4)-\mathcal{F}_3(-k_1,-k_2,-k_{12},k_{12},k_3,k_4)\bigg)
      \nonumber\\&&\quad\quad\quad\quad
      +6(\mathbf{k}_{12}\cdot\mathbf{k}_4)\bigg(\mathcal{F}_3(k_1,k_2,-k_{12},k_3,k_4,k_{12})-\mathcal{F}_3(-k_1,-k_2,-k_{12},k_3,k_4,k_{12})\bigg)
      \nonumber\\&&\quad\quad\quad\quad
      +3(\mathbf{k}_1\cdot\mathbf{k}_2)\bigg(\mathcal{F}_4(-k_{12},k_1,k_2,k_3,k_4,k_{12})-\mathcal{F}_4(-k_{12},-k_1,-k_2,k_3,k_4,k_{12})\bigg)
      \nonumber\\&&\quad\quad\quad\quad
      -6(\mathbf{k}_{12}\cdot\mathbf{k}_2)\bigg(\mathcal{F}_4(k_1,k_2,-k_{12},k_3,k_4,k_{12})-\mathcal{F}_4(-k_1,-k_2,-k_{12},k_3,k_4,k_{12})\bigg)
\Bigg)
\nonumber\\
&&-B^2
\Bigg(
      (\mathbf{k}_1\cdot\mathbf{k}_2)(\mathbf{k}_3\cdot\mathbf{k}_4)\bigg(\mathcal{F}_2(-k_{12},k_1,k_2,k_{12},k_3,k_4)-\mathcal{F}_2(-k_{12},-k_1,-k_2,k_{12},k_3,k_4)\bigg)
      \nonumber\\&&\quad\quad\quad\quad
      +2(\mathbf{k}_1\cdot\mathbf{k}_2)(\mathbf{k}_{12}\cdot\mathbf{k}_4)\bigg(\mathcal{F}_2(-k_{12},k_1,k_2,k_3,k_4,k_{12})-\mathcal{F}_2(-k_{12},-k_1,-k_2,k_3,k_4,k_{12})\bigg)
      \nonumber\\&&\quad\quad\quad\quad
      -2(\mathbf{k}_{12}\cdot\mathbf{k}_2)(\mathbf{k}_3\cdot\mathbf{k}_4)\bigg(\mathcal{F}_2(k_1,k_2,-k_{12},k_{12},k_3,k_4)-\mathcal{F}_2(-k_1,-k_2,-k_{12},k_{12},k_3,k_4)\bigg)
      \nonumber\\&&\quad\quad\quad\quad
      -4(\mathbf{k}_{12}\cdot\mathbf{k}_2)(\mathbf{k}_{12}\cdot\mathbf{k}_4)\bigg(\mathcal{F}_2(k_1,k_2,-k_{12},k_3,k_4,k_{12})-\mathcal{F}_2(-k_1,-k_2,-k_{12},k_3,k_4,k_{12})\bigg)
\Bigg)\Bigg]
\nonumber\\&&
\,\,
+23\,\mathrm{permutations\, of}\{k_1,k_2,k_3,k_4\},\label{SETrispectrumPHI}
\end{eqnarray}
where as before $\mathbf{K}=\mathbf{k}_1+\mathbf{k}_2+\mathbf{k}_3+\mathbf{k}_4$, $N=H/\sqrt{2P_{,X}c_s}$ and $k_{12}=|\mathbf{k}_{12}|=|\mathbf{k}_1+\mathbf{k}_2|$. Similarly, we define $k_{ab}$ as $k_{ab}=|\mathbf{k}_{ab}|=|\mathbf{k}_{a}+\mathbf{k}_{b}|$, where $\mathbf{k}_{a}$ and $\mathbf{k}_{b}$ represent any of the four momentum vectors $\mathbf{k}_{1}$, $\mathbf{k}_{2}$, $\mathbf{k}_{3}$ and $\mathbf{k}_{4}$. Momentum conservation implies $k_{12}=k_{34}$, $k_{13}=k_{24}$ and $k_{14}=k_{23}$.

Because there are only four different types of double time integrals over the mode functions, we define four $\mathcal{F}_i$ functions (with $i=1,\ldots,4$). Their explicit expressions are
\begin{eqnarray}
\mathcal{F}_1(k_1,k_2,k_3,k_4,k_5,k_6)&=&\int_{-\infty}^0d\eta a(\eta)\int_{-\infty}^\eta d\tilde\eta a(\tilde\eta)U^{*'}(\eta,k_1)U^{*'}(\eta,k_2)U^{*'}(\eta,k_3)U^{*'}(\tilde\eta,k_4)U^{*'}(\tilde\eta,k_5)U^{*'}(\tilde\eta,k_6)
\nonumber\\
&=&-4\frac{N^6c_s^6}{H^2}|k_1\cdots k_6|^\frac{1}{2}\frac{1}{\mathcal{A}^3\mathcal{C}^3}\left(1+3\frac{\mathcal{A}}{\mathcal{C}}+6\frac{\mathcal{A}^2}{\mathcal{C}^2}\right),
\end{eqnarray}
where $\mathcal{A}$ is defined by the sum of the last three arguments of the $\mathcal{F}_i$ functions as  $\mathcal{A}=k_4+k_5+k_6$ and $\mathcal{C}$ is defined by the sum of all the arguments as $\mathcal{C}=k_1+k_2+k_3+k_4+k_5+k_6$. The remaining functions are
\begin{eqnarray}
\mathcal{F}_2(k_1,k_2,k_3,k_4,k_5,k_6)&=&\int_{-\infty}^0d\eta a(\eta)\int_{-\infty}^\eta d\tilde\eta a(\tilde\eta)U^{*'}(\eta,k_1)U^{*}(\eta,k_2)U^{*}(\eta,k_3)U^{*'}(\tilde\eta,k_4)U^{*}(\tilde\eta,k_5)U^{*}(\tilde\eta,k_6)
\nonumber\\
&=&-\frac{N^6c_s^2}{H^2}\frac{|k_1k_4|^\frac{1}{2}}{|k_2k_3k_5k_6|^\frac{3}{2}}\frac{1}{\mathcal{A}\mathcal{C}}
\bigg[
      1+\frac{k_5+k_6}{\mathcal{A}}+2\frac{k_5k_6}{\mathcal{A}^2}
      \nonumber\\&&
      +\frac{1}{\mathcal{C}}  \left(k_2+k_3+k_5+k_6+\frac{1}{\mathcal{A}}\left(\left(k_2+k_3\right)\left(k_5+k_6\right)+2k_5k_6\right)+2\frac{k_5k_6\left(k_2+k_3\right)}{\mathcal{A}^2}\right)
      \nonumber\\&&
      +\frac{2}{\mathcal{C}^2}\bigg(k_5k_6+\left(k_2+k_3\right)\left(k_5+k_6\right)+k_2k_3
      \nonumber\\&&
      \qquad\quad+\frac{1}{\mathcal{A}}\left(k_2k_3\left(k_5+k_6\right)+2k_5k_6\left(k_2+k_3\right)\right)+2\frac{k_2k_3k_5k_6}{\mathcal{A}^2}\bigg)
      \nonumber\\&&
      +\frac{6}{\mathcal{C}^3}\left(k_2k_3\left(k_5+k_6\right)+k_5k_6\left(k_2+k_3\right)+2\frac{k_2k_3k_5k_6}{\mathcal{A}}\right)+24\frac{k_2k_3k_5k_6}{\mathcal{C}^4}
\bigg],
\end{eqnarray}
\begin{eqnarray}
\mathcal{F}_3(k_1,k_2,k_3,k_4,k_5,k_6)&=&\int_{-\infty}^0d\eta a(\eta)\int_{-\infty}^\eta d\tilde\eta a(\tilde\eta)U^{*'}(\eta,k_1)U^{*'}(\eta,k_2)U^{*'}(\eta,k_3)U^{*'}(\tilde\eta,k_4)U^{*}(\tilde\eta,k_5)U^{*}(\tilde\eta,k_6)
\nonumber\\
&=&2\frac{N^6c_s^4}{H^2}\frac{|k_1k_2k_3k_4|^\frac{1}{2}}{|k_5k_6|^\frac{3}{2}}\frac{1}{\mathcal{A}\mathcal{C}^3}
\left[
      1+\frac{k_5+k_6}{\mathcal{A}}+2\frac{k_5k_6}{\mathcal{A}^2}
      +\frac{3}{\mathcal{C}}\left(k_5+k_6+2\frac{k_5k_6}{\mathcal{A}}\right)
      +12\frac{k_5k_6}{\mathcal{C}^2}
\right],\nonumber\\
\end{eqnarray}
\begin{eqnarray}
\mathcal{F}_4(k_1,k_2,k_3,k_4,k_5,k_6)&=&\int_{-\infty}^0d\eta a(\eta)\int_{-\infty}^\eta d\tilde\eta a(\tilde\eta)U^{*'}(\eta,k_1)U^{*}(\eta,k_2)U^{*}(\eta,k_3)U^{*'}(\tilde\eta,k_4)U^{*'}(\tilde\eta,k_5)U^{*'}(\tilde\eta,k_6)
\nonumber\\
&=&2\frac{N^6c_s^4}{H^2}\frac{|k_1k_4k_5k_6|^\frac{1}{2}}{|k_2k_3|^\frac{3}{2}}\frac{1}{\mathcal{A}^3\mathcal{C}}
\bigg[1+\frac{\mathcal{A}}{\mathcal{C}}+\frac{\mathcal{A}^2}{\mathcal{C}^2}+\frac{k_2+k_3}{\mathcal{C}}+2\frac{\mathcal{A}\left(k_2+k_3\right)+k_2k_3}{\mathcal{C}^2}
\nonumber\\&&
\qquad\qquad\qquad\qquad\qquad\quad
+3\frac{\mathcal{A}}{\mathcal{C}^3}\left(\mathcal{A}\left(k_2+k_3\right)+2k_2k_3\right)+12k_2k_3\frac{\mathcal{A}^2}{\mathcal{C}^4}\bigg],
\end{eqnarray}
where the function $U(\eta,k)$ is a modified mode function given by
\begin{equation}
U(\eta,k)=N\frac{1}{|k|^\frac{3}{2}}\left(1+ikc_s\eta\right)e^{-ikc_s\eta}.
\end{equation}
If the sign of the argument $k$ of $U$ is positive then $U$ is equal to the mode function, if the sign is negative then $U$ equals the complex conjugate of the mode function.

The study of the momentum dependence of the trispectrum will be the
subject of the next section, but here we would like to make a remark. In
the configuration where all the momentum vectors have equal magnitude,
it is easy to see that the variable $\mathcal{C}$ defined previously and
that appears in some $\mathcal{F}_i$ functions in
(\ref{SETrispectrumPHI}) becomes zero. Because $\mathcal{C}$ appears in
the denominator of some fractions in the definitions of $\mathcal{F}_i$
one might naively think that the trispectrum diverges for these
configurations. This is not the case. In Appendix \ref{DIV}, we show that if one includes all the permutations of these terms the apparent divergences do not appear anymore and as expected the trispectrum is finite for these configurations.

As in the previous subsection, the four-point function of the curvature perturbation $\zeta$ is related with the four-point function of the field perturbation as
\begin{eqnarray}
\langle\Omega|\zeta(0,\mathbf{k}_1)\zeta(0,\mathbf{k}_2)\zeta(0,\mathbf{k}_3)\zeta(0,\mathbf{k}_4)|\Omega\rangle^{SE}=\frac{H^4}{\dot\phi^4}\langle\Omega|\delta\phi(0,\mathbf{k}_1)\delta\phi(0,\mathbf{k}_2)\delta\phi(0,\mathbf{k}_3)\delta\phi(0,\mathbf{k}_4)|\Omega\rangle^{SE}.
\end{eqnarray}
This constitutes one of the main results of this work.

The total trispectrum of $\zeta$ is the sum of the contributions coming from both of the diagrams of Fig. \ref{TrispectrumDiagrams} as
\begin{eqnarray}
\langle\Omega|\zeta(0,\mathbf{k}_1)\zeta(0,\mathbf{k}_2)\zeta(0,\mathbf{k}_3)\zeta(0,\mathbf{k}_4)|\Omega\rangle^{total}&=&\langle\Omega|\zeta(0,\mathbf{k}_1)\zeta(0,\mathbf{k}_2)\zeta(0,\mathbf{k}_3)\zeta(0,\mathbf{k}_4)|\Omega\rangle^{CI}
\nonumber\\&&+\langle\Omega|\zeta(0,\mathbf{k}_1)\zeta(0,\mathbf{k}_2)\zeta(0,\mathbf{k}_3)\zeta(0,\mathbf{k}_4)|\Omega\rangle^{SE}.
\label{TotalTriZeta}
\end{eqnarray}

From the previous result one can read the order of magnitude of the total trispectrum and it agrees with the discussion of subsection \ref{subsec:OM}. Moreover, one can also read the exact momentum dependence of the trispectrum. In the next section we shall study the shape of the trispectrum in more detail. In particular, we will consider the so-called equilateral configuration and we shall calculate the value of the non-linearity parameter $\tau_{NL}$.


\section{\label{sec:Shape}$\tau_{NL}$ and the shape of the trispectrum
in the equilateral configuration}

In this section, we will calculate the non-linearity parameter $\tau_{NL}$ in the limit where all the four momentum vectors have the same magnitude $k$. This is the so-called equilateral configuration used in previous works \cite{Seery:2006vu,Huang:2006eh}.
If we denote the angle between $\mathbf{k}_i$ and $\mathbf{k}_j$ by $\theta_{ij}$, then in this configuration, we have $\cos(\theta_{12})=\cos(\theta_{34})\equiv\cos(\theta_3)$, $\cos(\theta_{23})=\cos(\theta_{14})\equiv\cos(\theta_1)$, $\cos(\theta_{13})=\cos(\theta_{24})\equiv\cos(\theta_2)$ and $\cos(\theta_1)+\cos(\theta_2)+\cos(\theta_3)=-1$ due to momentum conservation.

The inflationary trispectrum is often parameterized \cite{Byrnes:2006vq,Seery:2006js,Seery:2008ax} by two non-linearity parameters $\tau_{NL}^{local}$ and $g_{NL}^{local}$ as
\begin{eqnarray}
\langle\zeta(\mathbf{k}_1)\zeta(\mathbf{k}_2)\zeta(\mathbf{k}_3)\zeta(\mathbf{k}_4)\rangle&=&\tau_{NL}^{local}\left(\tilde {\cal P}_\zeta(k_{13})\tilde {\cal P}_\zeta(k_{3})\tilde {\cal P}_\zeta(k_{4})+11\,\mathrm{permutations}\right)
\nonumber\\
&&+\frac{54}{25}g_{NL}^{local}\left(\tilde {\cal P}_\zeta(k_{2})\tilde {\cal P}_\zeta(k_{3})\tilde {\cal P}_\zeta(k_{4})+3\,\mathrm{permutations}\right),\label{localTrispectrum}
\end{eqnarray}
where the power spectrum $\tilde {\cal P}_\zeta(k)$ is given by $\langle\zeta(\mathbf{k}_1)\zeta(\mathbf{k}_2)\rangle=(2\pi)^3\delta^{(3)}(\mathbf{k}_1+\mathbf{k}_2)\tilde {\cal P}_\zeta(k_1)$ or $\tilde {\cal P}_\zeta(k)=H^2/(4\epsilon c_sk^3)$.
This shape of the trispectrum results from the most general local parametrization for the non-linearity of $\zeta$ as
\begin{equation}
\zeta(\mathbf{x})=\zeta_G(\mathbf{x})+\frac{1}{2}(\tau_{NL}^{local})^{1/2}\left(\zeta_G^2(\mathbf{x})-\overline{\zeta_G^2(\mathbf{x})}\right)+\frac{9}{25}g_{NL}^{local}\zeta_G^3(\mathbf{x}),
\label{localNG}
\end{equation}
where $\zeta_G(\mathbf{x})$ denotes the first order curvature perturbation (gaussian random variable), $\overline{\zeta_G^2(\mathbf{x})}$ denotes the average of $\zeta_G^2$. The trispectrum calculated in the present work results from quantum correlations around horizon crossing and does not admit a simple parametrization in terms of momentum independent non-linearity parameters as (\ref{localTrispectrum}).

In the next subsections, we will adopt a naive parametrization for the trispectrum as
\begin{eqnarray}
\tau_{NL}(\mathbf{k}_1,\mathbf{k}_2,\mathbf{k}_3,\mathbf{k}_4)&=&\left(\frac{4\epsilon c_s}{H^2}\right)^3T_\zeta(\mathbf{k}_1,\mathbf{k}_2,\mathbf{k}_3,\mathbf{k}_4) \Pi_{i=1}^4k_i^3\nonumber\\&&\!\!\!\!\!\times
\bigg[
      \left(k_1^3k_2^3+k_3^3k_4^3\right)\left(k_{13}^{-3}+k_{14}^{-3}\right)
      +\left(k_1^3k_4^3+k_2^3k_3^3\right)\left(k_{12}^{-3}+k_{13}^{-3}\right)
      +\left(k_1^3k_3^3+k_2^3k_4^3\right)\left(k_{12}^{-3}+k_{14}^{-3}\right)
\bigg]^{-1},\nonumber\\
\end{eqnarray}
where $T_\zeta$ is related to the trispectrum of $\zeta$ as
\begin{equation}
\langle\zeta(\mathbf{k}_1)\zeta(\mathbf{k}_2)\zeta(\mathbf{k}_3)\zeta(\mathbf{k}_4)\rangle=(2\pi)^3\delta^{(3)}(\mathbf{K})T_\zeta(\mathbf{k}_1,\mathbf{k}_2,\mathbf{k}_3,\mathbf{k}_4).
\end{equation}
In general, $\tau_{NL}$ defined by the previous expression will be shape dependent. For instance, this will be the case for the contact interaction trispectrum. Only for non-Gaussianity of the form
(\ref{localNG}) with $g_{NL}^{local}=0$ the parameter $\tau_{NL}$ will be a constant, i.e. shape independent.
If one fully specifies the momentum vectors configuration, then $\tau_{NL}$ becomes just a number. This seems to be the approach followed by several authors in the literature and in the next subsections we shall do the same
to obtain the order of magnitude of the trispectrum.  However, we should emphasize that the full trispectrum contains much more information than the one that can be described by just a number like $\tau_{NL}$ and that this parametrization might not be so adequate to describe the information encoded in a general trispectrum.

\subsection{\label{sec:generalmodel} $\tau_{NL}$: general model}

For the general model (\ref{action}) and for an angular configuration like $\cos(\theta_1)=\cos(\theta_2)=\cos(\theta_3)=-1/3$ the contact interaction trispectrum gives a non-linearity parameter as
\begin{equation}
{\tau_{NL}}^{CI}_{\beta_1}\sim-0.14\frac{H^2c_s^2\epsilon}{P_{,X}^2}\beta_1, \quad
{\tau_{NL}}^{CI}_{\beta_2}\sim0.21\frac{H^2\epsilon}{P_{,X}^2}\beta_2, \quad
{\tau_{NL}}^{CI}_{\beta_3}\sim-0.55\frac{H^2\epsilon}{c_s^2P_{,X}^2}\beta_3,\label{tauNLCI}
\end{equation}
where the different values of $\tau_{NL}$ correspond to the three terms in the contact interaction trispectrum Eq. (\ref{CITrispectrumPHI}).

The scalar exchange trispectrum gives values for $\tau_{NL}$ as
\begin{equation}
{\tau_{NL}}^{SE}_{A^2}\sim0.52\frac{H^2c_s^4\epsilon}{P_{,X}^3}A^2, \quad
{\tau_{NL}}^{SE}_{AB}\sim-1.75\frac{H^2c_s^2\epsilon}{P_{,X}^3}AB, \quad {\tau_{NL}}^{SE}_{B^2}\sim2.51\frac{H^2\epsilon}{P_{,X}^3}B^2,
\label{tauNLSEeq}
\end{equation}
where ${\tau_{NL}}^{SE}_{A^2}$, ${\tau_{NL}}^{SE}_{B^2}$ and ${\tau_{NL}}^{SE}_{AB}$ correspond to the values of $\tau_{NL}$ calculated from the terms in (\ref{SETrispectrumPHI}) proportional to $A^2$, $B^2$ and $AB$ respectively.


\subsection{\label{subsec:Shape}The shape of the trispectrum for the
DBI-inflation model in the equilateral configuration}

In this subsection, we shall study the shape of the trispectrum for the DBI-inflation model in the equilateral configuration. We will assume that the trispectrum is maximized in the equilateral configuration and we will study the dependence of the maximum on the two remaining degrees of freedom, the angles $\theta_1$ and $\theta_2$\footnote{Contrary to what happens in the case of the bispectrum, the equilateral configuration conditions do not fix all degrees of freedom (dof) required to describe the shape of the trispectrum and we are left with two angular dof \cite{Seery:2006vu}}.
We will show that the trispectrum is maximized in the configuration where the angles between the four different momentum vector are equal. Although we do not prove that the trispectrum has the maximum value for the equilateral configuration, we present some evidence indicating that this might be the case in Appendix \ref{MAX}.

\begin{figure}[t]
\centerline{
\includegraphics[width=6.5cm]{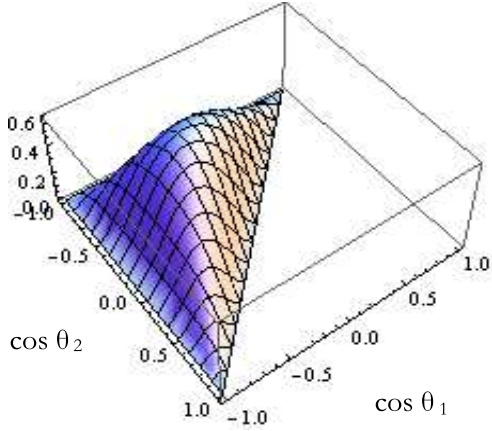}}
\caption{Plot of the non-linearity parameter $\tau_{NL}^{SE}$ calculated from the scalar exchange trispectrum. The maximum amplitude occurs for $\cos\theta_1=\cos\theta_2=-1/3$ and its value was rescaled by $c_s^4$.}
\label{tauNLSE}
\end{figure}

In the equilateral configuration, the contact interaction trispectrum is constant and only the scalar exchange trispectrum depends on $\theta_1$ and $\theta_2$. The plot of the shape of the scalar exchange trispectrum can be found in Fig. \ref{SEtrispectrum}. The exact analytical expression can be found in Appendix \ref{TDBI}. The shape of the trispectrum has a maximum for $\cos\theta_1=\cos\theta_2=\cos\theta_3=-1/3$ as it can be better seen in the density plot of the trispectrum in the r.h.s panel of Fig. \ref{SEtrispectrum}. In Fig. \ref{tauNLSE}, we plot the non-linearity parameter $\tau_{NL}^{SE}$ calculated from the scalar exchange trispectrum. The maximum amplitude occurs for $\cos\theta_1=\cos\theta_2=\cos\theta_3=-1/3$.

When $\cos\theta_1=-1$ or $\cos\theta_2=-1$ or $\cos\theta_3=-1$ (on the diagonal line in the plot of Fig. \ref{tauNLSE}, when $\cos\theta_2=-\cos\theta_1$) the non-linearity parameter is zero. This result can be nicely understood using the counter-collinear limit of the scalar exchange trispectrum \cite{Seery:2008ax}.
According to \cite{Seery:2008ax}, in the limit where the momentum of the scalar mode that is exchanged goes to zero, one can find a simple relation between the scalar exchange trispectrum and the power spectrum, in a similar way to Maldacena's consistency relations \cite{Maldacena:2002vr} (see also \cite{Creminelli:2004yq,Huang:2006eh,Cheung:2007sv}) since one can treat this mode as a background.
 Summarizing their result, let us suppose that the momentum of the exchanged particle is $k_{12}$ and that $k_{12}\ll k_1\approx k_2,k_3\approx k_4$. Then the mode associated with this scalar particle will cross the horizon much before the other $k_i$ modes, where $i=1,\ldots,4$, and its only effect is to rescale the spatial background where the $k_i$ modes exist. After some algebra, Ref. \cite{Seery:2008ax} found that the scalar exchange trispectrum obeys the following relation
\begin{equation}
\langle\zeta(\mathbf{k}_1)\zeta(\mathbf{k}_2)\zeta(\mathbf{k}_3)\zeta(\mathbf{k}_4)\rangle ^{SE} \rightarrow (2\pi)^3\delta^{(3)}(\mathbf{K})(n_s-1)^2\tilde {\cal P}_\zeta(k_{12})\tilde {\cal P}_\zeta(k_{1})\tilde {\cal P}_\zeta(k_{3}),
\label{consistencyrel}
\end{equation}
in the limit $\mathbf{k}_{12}\rightarrow0$. The previous equation implies that in the counter-collinear limit, the scalar exchange non-linearity parameter $\tau_{NL}^{SE}$ is of order $\epsilon^2$ (at leading order in our approximations this is equivalent to say that $\tau_{NL}^{SE}$ should vanish). We have verified that $\tau_{NL}^{SE}$ calculated from (\ref{SETrispectrumPHI}) does indeed vanish in the limit $k_{ab}\rightarrow0$. This provides a consistency check on (\ref{SETrispectrumPHI}). If we note that in the equilateral configuration $k_{14}=k_{23}=\sqrt{2}k\sqrt{1+\cos\theta_1}$, $k_{24}=k_{13}=\sqrt{2}k\sqrt{1+\cos\theta_2}$ and $k_{34}=k_{12}=\sqrt{2}k\sqrt{1+\cos\theta_3}$, then it is easy to understand why in the plot of Fig. \ref{tauNLSE}, $\tau_{NL}^{SE}$ vanishes for $\cos\theta_j=-1$ with $j=1,2,3$. This is because we are in the domain of validity of the counter-collinear limit relation.

\begin{figure}[t]
\centerline{
\includegraphics[width=12cm]{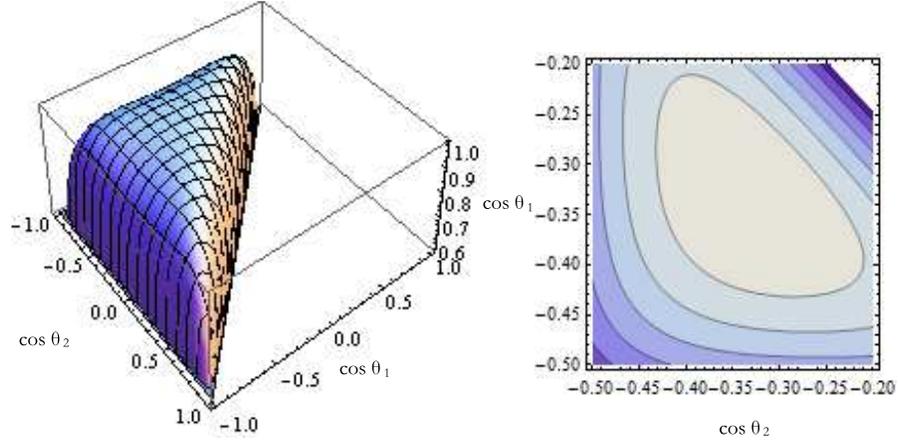}}
\caption{{\it Left}: The shape of the scalar exchange trispectrum as a function of the variables
$\cos \theta_1$ and $\cos \theta_2$.
The maximum amplitude has been normalized to unity. {\it Right}: Density plot of the shape of
the scalar exchange trispectrum as a function of the variables $\cos\theta_1$ and $\cos\theta_2$.
The maximum amplitude occurs for $\cos\theta_1=\cos\theta_2=-1/3$.}
\label{SEtrispectrum}
\end{figure}


\subsection{\label{subsec:DBImodel} $\tau_{NL}$: DBI-inflation}

In order to give numerical results for $\tau_{NL}$, one has to choose a particular model which will determine the parameters $\beta_i$, $A$ and $B$. In this subsection, we particularize the results (\ref{tauNLCI}) and (\ref{tauNLSEeq}) for
the DBI-inflation model at leading order in the small sound speed.

The total $\tau_{NL}$ coming from the scalar exchange trispectrum is
\begin{equation}
{\tau_{NL}}^{SE}_{DBI}={\tau_{NL}}^{SE}_{A^2}+{\tau_{NL}}^{SE}_{B^2}+{\tau_{NL}}^{SE}_{AB}\sim\frac{0.60}{c_s^4}.
\end{equation}

This result should be summed with $\tau_{NL}$ coming from the contact interaction trispectrum \cite{Huang:2006eh}
\begin{equation}
{\tau_{NL}}^{CI}_{DBI}={\tau_{NL}}^{CI}_{\beta_1}\sim-\frac{0.04}{c_s^4}.
\end{equation}
For this particular configuration the magnitude of ${\tau_{NL}}^{SE}_{DBI}$ is more than one order of magnitude higher than ${\tau_{NL}}^{CI}_{DBI}$.
We conclude that the total non-linearity parameter $\tau_{NL}$ for DBI-inflation in the equilateral configuration with equal angles between the momentum vectors is
\begin{equation}
{\tau_{NL}}^{equi}_{DBI}\sim\frac{0.56}{c_s^4}.
\end{equation}


\section{\label{sec:conclusion}Conclusion}

In this work we have computed the connected four-point function of the primordial curvature perturbation at leading order in the slow-roll expansion and in the small sound speed limit for a generic model of kinetically driven inflation. This four-point correlation is coming from either via the exchange of a scalar mode or by a four-leg contact interaction at a point. We showed that in general the scalar exchange correlation is as important as the contact interaction correlation previously calculated in the literature. So the final answer to be compared with observational data should include the sum of these two contributions.

We have studied the shape dependence of the trispectrum for the DBI-inflation model in the equilateral configurations. The equilateral limit conditions do not fix completely the configuration of the four momentum vectors and we are left with two remaining angular degrees of freedom. We have shown that the trispectrum is maximized when the angles between the four different momentum vectors are equal to $\arccos(-1/3)\approx109.5^{\circ}$. We found that for this ``maximal" equilateral configuration the non-linearity parameter for DBI-inflation ${\tau_{NL}}_{DBI}^{equi}$ is about one order of magnitude larger than the existing result in the literature that only takes into account the contact interaction diagram. Our final result for the non-linearity parameter for the DBI-inflation model is
${\tau_{NL}}_{DBI}^{equi}\sim0.56/c_s^4$.

We did not prove that the maximum of the trispectrum occurs for the equilateral configuration. But we gave some evidence in Appendix B. In general there are five parameters to characterize the shape of the trispectrum.
It would be desirable to explore, possibly numerically, all configurations to show that
the maximum of $\tau_{NL}$ lies in the configuration we found. Also it is necessary to construct an
estimator to extract information from CMB anisotropies. We will leave these issues for future investigations.

\emph{Note added}: On the day this work was submitted, the paper \cite{Chen:2009bc} appeared in the arXiv, which contains some similar results.

\begin{acknowledgments}
We would like to thank Filippo Vernizzi for interesting discussions. We thank Takashi Hiramatsu for pointing out some typos in an early version of the manuscript.
FA and SM are supported by the Japanese Society for the Promotion of Science (JSPS).
KK is supported by RCUK, STFC and ERC.
The authors thank the Yukawa Institute for Theoretical Physics (YITP) at
 Kyoto University. Discussions during the YITP workshop YITP-W-09-01 on
``Non-linear cosmological perturbations" were useful to complete this
 work. KK's visit to the Yukawa Institute was supported by the Royal Society
international joint project grant.
  We also thank the Institute for the Physics and Mathematics of
 the Universe (IPMU) for organizing the workshop entitled ``Focus week on
 non-Gaussianities in the sky" during which we had several important
 discussions about this work. SM is grateful to the ICG,
Portsmouth for their hospitality when this work was
completed.
\end{acknowledgments}

\appendix
\section{\label{DIV}Canceling the apparent divergences of the trispectrum in the equilateral limit}

In the configuration where all the momentum vectors have equal magnitude, i.e. the equilateral configuration, the variable $\mathcal{C}$ defined as the sum of the arguments of the $\mathcal{F}_i$ functions vanishes for the terms in (\ref{SETrispectrumPHI}) that contain a $\mathcal{F}_i$ function with three negative arguments. Because $\mathcal{C}$ appears in the denominator of some fractions in the definitions of $\mathcal{F}_i$ one might naively think that the trispectrum (\ref{SETrispectrumPHI}) diverges for these configurations.
We will now show that when considering all the 23 permutations in (\ref{SETrispectrumPHI}), the powers of $\mathcal{C}$ that appear in the denominator in the different terms cancel out and we are left with expressions that are clearly finite in the limit $\mathcal{C}$ going to zero.

The proof is rather simple if one notes that the apparently divergent terms ($\mathcal{F}_i$ functions with three negative arguments) can be gathered in certain combinations that are clearly finite in the equilateral limit.

For example, if we take all the divergent terms containing $\mathcal{F}_1$, we find pairs of terms like $\mathcal{F}_1(-k_a,-k_b,-k_{ab},k_c,k_d,k_{ab})+\mathcal{F}_1(-k_c,-k_d,-k_{ab},k_a,k_b,k_{ab})$,
where the only difference from the first term to the second is that $k_a,k_b$ changed position with $k_c,k_d$, where $k_a,k_b,k_c,k_d$ can represent any of the momentum vectors $k_1,k_2,k_3,k_4$ and $k_{ab}$ is $k_{ab}=|\mathbf{k}_{ab}|=|\mathbf{k}_a+\mathbf{k}_b|$.
To complete the proof, one just needs to show that this combination is finite in the equilateral limit. After some algebra one can show that
\begin{eqnarray}
\mathcal{F}_1(-k_a,-k_b,-k_{ab},k_c,k_d,k_{ab})+\mathcal{F}_1(-k_c,-k_d,-k_{ab},k_a,k_b,k_{ab})=4\frac{N^6c_s^6}{H^2}|k_ak_bk_ck_dk_{ab}^2|^{\frac{1}{2}}\frac{1}{\mathcal{A}_1^3\mathcal{A}_2^3},
\end{eqnarray}
where $\mathcal{A}_1=-(k_{ab}+k_c+k_d)$ and $\mathcal{A}_2=-(k_{ab}+k_a+k_b)$. The r.h.s. of the previous equation is clearly finite in the limit $\mathcal{C}\rightarrow 0$.

The proof of finiteness for the remaining divergent terms containing $\mathcal{F}_3$ and $\mathcal{F}_4$ in (\ref{SETrispectrumPHI}) is completely analogous to the proof just described but now one has to use the following combinations of terms.
\begin{eqnarray}
\mathcal{F}_3(-k_a,-k_b,-k_{ab},k_{ab},k_c,k_d)&+&\mathcal{F}_4(-k_{ab},-k_c,-k_d,k_a,k_b,k_{ab})
\nonumber\\
&=&2\frac{N^6c_s^4}{H^2}\frac{|k_ak_bk_{ab}^2|^{\frac{1}{2}}}{|k_ck_d|^{\frac{3}{2}}}\bigg[-\frac{1}{\mathcal{A}_1\mathcal{A}_2^3}-2\frac{k_ck_d}{\mathcal{A}_1^3\mathcal{A}_2^3}+\frac{k_c+k_d}{\mathcal{A}_1^2\mathcal{A}_2^3}\bigg],
\end{eqnarray}
\begin{eqnarray}
\mathcal{F}_3(-k_a,-k_b,-k_{ab},k_c,k_d,k_{ab})&+&\mathcal{F}_4(-k_c,-k_d,-k_{ab},k_a,k_b,k_{ab})
\nonumber\\
&=&2\frac{N^6c_s^4}{H^2}\frac{|k_ak_bk_ck_{ab}|^{\frac{1}{2}}}{|k_dk_{ab}|^{\frac{3}{2}}}\bigg[-\frac{1}{\mathcal{A}_1\mathcal{A}_2^3}-2\frac{k_dk_{ab}}{\mathcal{A}_1^3\mathcal{A}_2^3}+\frac{k_d+k_{ab}}{\mathcal{A}_1^2\mathcal{A}_2^3}\bigg].
\end{eqnarray}
Finally, all divergent terms containing $\mathcal{F}_2$ can be completely written in terms of the following combinations
\begin{eqnarray}
\mathcal{F}_2(-k_{ab}&&\!\!\!\!\!\!\!\!\!\!\!\!,-k_a,-k_b,k_{ab},k_c,k_d)+\mathcal{F}_2(-k_{ab},-k_c,-k_d,k_{ab},k_a,k_b)
\nonumber\\
&=&\frac{N^6c_s^2}{H^2}\frac{|k_{ab}|}{|k_ak_bk_ck_d|^\frac{3}{2}}
\frac{
      \left[
            2\mathcal{A}_2^2+\mathcal{A}_2\left(k_{ab}-2k_a\right)-2k_a\left(k_a+k_{ab}\right)
      \right]
      \left[
            2\mathcal{A}_1^2+\mathcal{A}_1\left(k_{ab}-2k_d\right)-2k_d\left(k_d+k_{ab}\right)
      \right]
     }{\mathcal{A}_1^3\mathcal{A}_2^3}
      ,\nonumber\\
\end{eqnarray}
\begin{eqnarray}
\mathcal{F}_2(-k_{ab}&&\!\!\!\!\!\!\!\!\!\!\!\!,-k_a,-k_b,k_c,k_d,k_{ab})+\mathcal{F}_2(-k_c,-k_d,-k_{ab},k_{ab},k_a,k_b)
\nonumber\\
&=&\frac{N^6c_s^2}{H^2}\frac{|k_ck_{ab}|^\frac{1}{2}}{|k_ak_bk_dk_{ab}|^\frac{3}{2}}
\frac{
      \left[
            2\mathcal{A}_2^2+\mathcal{A}_2\left(k_{ab}-2k_a\right)-2k_a\left(k_a+k_{ab}\right)
      \right]
      \left[
            \mathcal{A}_1^2-\mathcal{A}_1\left(k_d+k_{ab}\right)+2k_dk_{ab}
      \right]
     }{\mathcal{A}_1^3\mathcal{A}_2^3}
      ,
\end{eqnarray}
\begin{eqnarray}
\mathcal{F}_2(-k_b&&\!\!\!\!\!\!\!\!\!\!\!\!,-k_a,-k_{ab},k_d,k_c,k_{ab})+\mathcal{F}_2(-k_d,-k_c,-k_{ab},k_b,k_a,k_{ab})
\nonumber\\
&=&\frac{N^6c_s^2}{H^2}\frac{|k_bk_d|^\frac{1}{2}}{|k_ak_ck_{ab}^2|^\frac{3}{2}}
\frac{
      \left[
            2\mathcal{A}_1^2+\mathcal{A}_1\left(k_d-2k_{ab}\right)-2k_{ab}\left(k_d+k_{ab}\right)
      \right]
      \left[
            \mathcal{A}_2^2-\mathcal{A}_2\left(k_a+k_{ab}\right)+2k_ak_{ab}
      \right]
     }{\mathcal{A}_1^3\mathcal{A}_2^3}
      .
\end{eqnarray}
All of these combinations are finite in the equilateral limit and all apparently divergent terms in the scalar exchange trispectrum (\ref{SETrispectrumPHI}) can be written in terms of these combinations. This implies that the trispectrum (\ref{SETrispectrumPHI}) is also finite in this limit as expected.

\section{\label{MAX}Where does the maximum of the trispectrum for the DBI-inflation model lie?}

It seems difficult to pursue an analytical answer for this question. The main reason is the fact that we need five degrees of freedom to parameterize the shape of the trispectrum and the expression for the trispectrum, Eq. (\ref{TotalTriZeta}), is fairly complicated. Numerical searches for the maximum in the full parameter space are possible even with such a large parameter space to explore. We leave this for future work.

In this Appendix, we will relax the condition of equilateral configuration, allowing for two of the momentum vectors to have a different magnitude from the other two and we will show that the maximum that we found in subsection \ref{subsec:Shape} is still a maximum of $\tau_{NL}$. This provides some evidence that this particular configuration maximizes $\tau_{NL}$.

Before we show the results, it is interesting to consider analytically
certain limiting configurations such as the squeezed limit or in the
case of the scalar exchange diagram the counter-collinear limit. These
limits give us other indications about where the maximum of the trispectrum might lie. Also, because these limits can be obtained using different methods from the one used in this work, they provide consistency checks with our final answers.

As first pointed out by Maldacena \cite{Maldacena:2002vr} and Seery \emph{et al.} \cite{Seery:2006vu}, all higher order correlators of $\zeta$ in single field inflation obey consistency relations in the squeezed limit, i.e. when one of the momentum vector is very small. Let us consider the limit $\mathbf{k}_1\rightarrow0$, which implies that the corresponding mode $\zeta(\mathbf{k}_1)$ leaves the horizon well before all other modes
leave the horizon. By the time the remaining modes exit the horizon, $\zeta(\mathbf{k}_1)$ will be frozen as a super-horizon mode and its only effect is to deform the background.
Following the algebra with some more detail, one finds the following relation for the trispectrum \cite{Seery:2006vu}
\begin{equation}
\langle\zeta(\mathbf{k}_1)\zeta(\mathbf{k}_2)\zeta(\mathbf{k}_3)\zeta(\mathbf{k}_4)\rangle \rightarrow-\tilde {\cal P}_\zeta(k_1)\frac{d}{Hdt}\langle\zeta(\mathbf{k}_2)\zeta(\mathbf{k}_3)\zeta(\mathbf{k}_4)\rangle,
\label{consistencyrel2}
\end{equation}
in the limit $\mathbf{k}_1\rightarrow0$. Using the fact that the bispectrum is $B_\zeta\sim c_s^{-2}{\cal \tilde P}_\zeta^2$ it follows that in the squeezed limit according to (\ref{consistencyrel2}) the trispectrum is of $\mathcal{O}(\epsilon c_s^{-2}{\cal \tilde P}_\zeta^3)$. Comparing this result with the full trispectrum, Eq. (\ref{TotalTriZeta}), which is of order $\epsilon^{-3}$, one finds that the total non-linearity parameter $\tau_{NL}$ calculated from (\ref{TotalTriZeta}) should be at most of order $\epsilon c_s^{-2}$ (i.e. under our approximation it should vanish) when any of the momentum vectors goes to zero. We verified that this is the case for (\ref{TotalTriZeta})\footnote{Strictly speaking, Eq. (\ref{TotalTriZeta}) is only valid when all momentum vectors have similar magnitude as only in this case we can treat, for instance, $H$ as a constant. However the momentum dependence of (\ref{TotalTriZeta}) is such that $\tau_{NL}^{Total}$ goes to zero in the squeezed limit and this is consistent with (\ref{consistencyrel2})}.
This would mean that the maximum of $\tau_{NL}$ might lie near the equilateral configuration.

\begin{figure}[t]
\centerline{
\includegraphics[width=14cm]{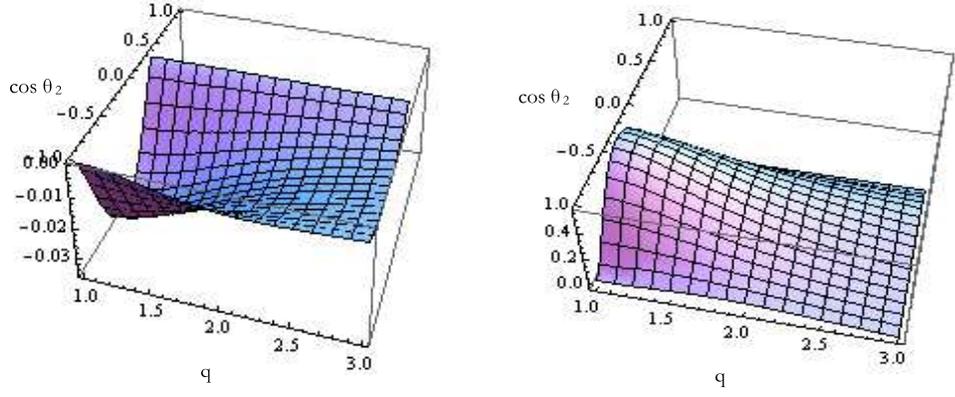}}
\caption{Plots of $\tau_{NL}$ coming from the contact interaction
 diagram and the total $\tau_{NL}$ as functions of the momentum
 amplitude $q$ and $\cos \theta_2$. The values were rescaled by $c_s^4$}
\label{fig:tauNLnonequi}
\end{figure}

As an example, we considered configurations where the two momentum vectors $\mathbf{k}_3$ and $\mathbf{k}_4$ have a different magnitude $q$ from the magnitude of $\mathbf{k}_1$ and $\mathbf{k}_2$ that we normalized to unity. Then in this configuration, we have $\cos(\theta_{34}) \equiv \cos(\theta_3)
, \cos(\theta_{23})=\cos(\theta_{14}) \equiv \cos(\theta_1), \cos(\theta_{13})=\cos(\theta_{24}) \equiv \cos(\theta_2), \cos(\theta_{12})= q^2 (1+\cos(\theta_3))-1$.
The momentum conservation implies $\cos\theta_3=-(1/q)(\cos\theta_1+\cos\theta_2)-1$.
We also allow one of the angular dof to vary freely, the other angular dof $\theta_1$ is fixed by $\cos\theta_1=-1/3$ to the value that gives us the maximum that we found in subsection \ref{subsec:Shape}. In Fig. \ref{fig:tauNLnonequi}, we plot the non-linearity parameter $\tau_{NL}$ for the DBI-inflation model.
From the momentum conservation, $\cos(\theta_2)$ is restricted to $\cos(\theta_2) \leq 1/3$. Also we restricted $q$ as $q \geq 1$ as the configurations with $q<1$ can be related to those with $q \geq 1$ by rescaling $k_1$ and $k_2$. The left panel shows the contribution from the contact interaction.
In the right panel, we plot the total $\tau_{NL}$. As can be seen in the r.h.s panel of Fig. \ref{fig:tauNLnonequi}, the maximum of $\tau_{NL}^{Total}$ occurs exactly for the equilateral configuration when all the angles $\theta_1$, $\theta_2$ and $\theta_3$ are equal. This provides evidence that the maximum of the non-linearity parameter would happen for this configuration.

\section{\label{TDBI}The DBI scalar exchange trispectrum in the equilateral limit}

In this Appendix, we present the scalar exchange trispectrum of the primordial curvature perturbation $\zeta$ for the DBI-inflation model in the equilateral configuration. These analytical expressions can be derived from the general expression for the scalar exchange trispectrum, Eq. (\ref{SETrispectrumPHI}). We will use these simpler equations to calculate the non-linearity parameter $\tau_{NL}$ in subsection \ref{subsec:DBImodel} and to study the shape dependence of the trispectrum in subsection \ref{subsec:Shape}.

At leading order in the small sound speed and in the slow-roll expansion, the scalar exchange trispectrum for the DBI inflationary model is
\begin{eqnarray}
{\langle\Omega|\zeta(0,\mathbf{k}_1)\zeta(0,\mathbf{k}_2)\zeta(0,\mathbf{k}_3)\zeta(0,\mathbf{k}_4)|\Omega\rangle^{SE}}&=&
{\langle\Omega|\zeta(0,\mathbf{k}_1)\zeta(0,\mathbf{k}_2)\zeta(0,\mathbf{k}_3)\zeta(0,\mathbf{k}_4)|\Omega\rangle^{SE}}_{A^2}
\nonumber\\
&&+{\langle\Omega|\zeta(0,\mathbf{k}_1)\zeta(0,\mathbf{k}_2)\zeta(0,\mathbf{k}_3)\zeta(0,\mathbf{k}_4)|\Omega\rangle^{SE}}_{B^2}
\nonumber\\
&&+{\langle\Omega|\zeta(0,\mathbf{k}_1)\zeta(0,\mathbf{k}_2)\zeta(0,\mathbf{k}_3)\zeta(0,\mathbf{k}_4)|\Omega\rangle^{SE}}_{AB},
\end{eqnarray}
where the three different contributions come from the terms in (\ref{SETrispectrumPHI}) that are proportional to $A^2$, $B^2$ and $AB$ respectively.

Explicitly, they are given by
\begin{eqnarray}
{\langle\Omega|\zeta(0,\mathbf{k}_1)\zeta(0,\mathbf{k}_2)\zeta(0,\mathbf{k}_3)\zeta(0,\mathbf{k}_4)|\Omega\rangle^{SE}}_{A^2}&=&(2\pi)^3\delta^{(3)}(\mathbf{K})
\frac{1}{c_s^4}\mathcal{D}_1(k_{12}/k)\tilde {\cal P}_\zeta^2(k)\tilde {\cal P}_\zeta(k_{12}) +2\,\mathrm{permutations},
\end{eqnarray}
\begin{eqnarray}
{\langle\Omega|\zeta(0,\mathbf{k}_1)\zeta(0,\mathbf{k}_2)\zeta(0,\mathbf{k}_3)\zeta(0,\mathbf{k}_4)|\Omega\rangle^{SE}}_{AB}&=&(2\pi)^3\delta^{(3)}(\mathbf{K})\frac{1}{c_s^4}
\mathcal{D}_2(k_{12}/k)
\tilde {\cal P}_\zeta^2(k)\tilde {\cal P}_\zeta(k_{12})
+2\,\mathrm{permutations},
\end{eqnarray}
\begin{eqnarray}
{\langle\Omega|\zeta(0,\mathbf{k}_1)\zeta(0,\mathbf{k}_2)\zeta(0,\mathbf{k}_3)\zeta(0,\mathbf{k}_4)|\Omega\rangle^{SE}}_{B^2}&=&(2\pi)^3\delta^{(3)}(\mathbf{K})\frac{1}{c_s^4}
\mathcal{D}_3(k_{12}/k)
\tilde {\cal P}_\zeta^2(k)\tilde {\cal P}_\zeta(k_{12}) +2\,\mathrm{permutations},
\end{eqnarray}
where we have defined three $\mathcal{D}_i(x)$ functions, with $i=1,2,3$, as
\begin{eqnarray}
\mathcal{D}_1(x)&=&\frac{9 x^4(3x^5+36x^4+176x^3+432x^2+528x+512)}{2^7 (x+2)^6},\nonumber\\
\mathcal{D}_2(x)&=&\frac{-3x^4(13x^7+156x^6+748x^5+1696x^4+1072x^3-4160x^2-11712x-8192)}{2^7(x+2)^6},\nonumber\\
\mathcal{D}_3(x)&=&\frac{x^4}{2^9 (x+2)^6}\times
\bigl(103x^9+1236x^8+5276x^7+5632x^6-27052x^5-95248x^4
-69376x^3
\nonumber\\
&&+149056x^2+281792x+131072\bigr),
\end{eqnarray}
where $k_{12}=k\sqrt{2(1+\cos\theta_3)}$, $k_{13}=k\sqrt{2(1+\cos\theta_2)}$ and $k_{14}=k\sqrt{2(1+\cos\theta_1)}$. $k$ denotes the common amplitude of all four momentum vectors $\mathbf{k}_j$, with $j=1,\ldots,4$ and the rescaled power spectrum is $\tilde {\cal P}_\zeta(k)=H^2/(4\epsilon c_sk^3)$ that should be evaluated at horizon crossing.



\end{document}